\documentclass[sigconf,nonacm,screen]{acmart}

\usepackage{booktabs}
\usepackage{array}
\usepackage{balance}
\usepackage{placeins}
\usepackage{xcolor}
\usepackage{enumitem}

\newif\ifcomments
\commentsfalse
\ifcomments
    \providecommand{\alvin}[1]{{\color{brown}{alvin: #1 }}}
    \providecommand{\ck}[1]{\textcolor{orange}{\footnotesize Mick: #1}}
\else
    \newcommand{\alvin}[1]{}
    \newcommand{\ck}[1]{}
\fi

\renewcommand\footnotetextcopyrightpermission[1]{}

\setlist[itemize]{leftmargin=*, topsep=0.5em}
\setlist[enumerate]{leftmargin=*, topsep=0.5em}

\renewcommand{\sectionautorefname}{\S\kern-1.5pt}
\renewcommand{\subsectionautorefname}{\S\kern-1.5pt}
\renewcommand{\subsubsectionautorefname}{\S\kern-1.5pt}

\begin{document}

\title{Concord: A Video Relational Algebra for Cross-Modal Query Optimization}

\newcommand{\berkeley}{%
  \affiliation{%
    \institution{UC Berkeley}%
    \city{Berkeley}%
    \state{California}%
    \country{USA}%
  }%
}

\settopmatter{authorsperrow=4}
\makeatletter
\author@bx@sep=1pt\relax %
\makeatother

\author{Sultan Muratbek}
\authornote{Equal contribution.}
\orcid{0009-0001-3735-0499}
\berkeley

\author{Charisse Ivana Yeung}
\authornotemark[1]
\orcid{0009-0006-5812-1165}
\berkeley

\author{Chanwut Kittivorawong}
\orcid{0000-0002-2884-2221}
\berkeley

\author{Alvin Cheung}
\orcid{0000-0001-6261-6263}
\berkeley

\begin{abstract}
Semantic video queries let users embed natural language prompts and use multimodal large language models (MLLMs) to
interpret the video. Such queries are increasingly popular for querying video data. However, their expressiveness comes at a steep cost: an MLLM may process hours of media to return only seconds of relevant output, making naive execution slow, expensive, and inaccurate.

We propose Concord, a system for expressing and optimizing semantic video
queries. We makes three contributions. First, we introduce \emph{Video
Relational Algebra} (VRA), a nested algebra over videos, transcripts,
frames, and object tracks that
captures common semantic video operations.
Second, we derive a set of approximate optimizations that rewrite VRA queries to reduce MLLM usage while improving result quality.
For narrated video, Concord either processes transcripts instead of video or
uses them to identify video clips for MLLM processing.
For cross-camera queries without narration, detection and tracking replace a
whole-video MLLM join with a track-level relational join.

Third, we evaluate Concord on real-world videos. Across 4.59 hours of soccer
broadcasts and 3.92 hours of lectures, transcript-to-video queries send only
5.32\% and 2.47\% of source-video duration to the MLLM and reduce MLLM cost by
up to 87\%. In two five-second highway clips with 18 manually adjudicated
cross-camera vehicles, a Detect-Track-Join query improves F1 from .364 to .813 while making no MLLM calls.
See our project at \href{https://concord-db.github.io}{concord-db.github.io}.
\end{abstract}

\maketitle

\section{Introduction}
\label{sec:intro}

Multimodal large language models (MLLMs) let users ask semantic questions over
videos: when a goal is scored in a soccer match, when an experiment visibly succeeds in a  lab demo, or which
vehicle appears across multiple camera views of the same highway. 
Naive execution sends complete, temporally redundant videos to an MLLM, incurring high cost and latency when relevant evidence is sparse. 
Furthermore, Video contains source-aligned modalities such as frames, 
audio, transcripts, and metadata. Many videos, including those on YouTube, already provide narration or transcripts.
In narrated videos, transcripts offer a cheaper access path: they may answer a query directly or identify candidate source-time intervals for video inference.
For example, a commentator announces a goal or a lecturer introduces a demo
before it is visible.
When the transcript lacks sufficient evidence to answer the query directly, it
can still identify \emph{candidate intervals}, source-time regions likely to
contain an answer.
The MLLM can then process these short regions instead of the complete video.

Not all videos have useful text channels. For cross-camera vehicle association, a direct implementation sends the synchronized videos to an MLLM, which must discover vehicles, maintain their identities within each video, and associate them across camera views. Concord instead decomposes this semantic operation into a \emph{Detect--Track--Join} query that derives entity tracks from pixels and joins them across cameras. This alternative changes the query's \emph{processing granularity} from whole videos to entity-level records.

Current video analytics systems optimize video processing through sampling, proxies, indexes, materialized views,
and relational hints~\cite{noscope,tasti,viva,seiden}, while
semantic data systems optimize programs whose operators may invoke foundation
models~\cite{palimpzest,docetl,thalamusdb}. What is missing is a compact
algebra in which ordinary relational operators compose with media-specific
representations.
Such an algebra expresses video-only inference,
transcript substitution, transcript-to-video pushdown, and track-level joins
as alternative queries with the same intent.

Towards that goal, we present \emph{Concord}, a declarative system for multimodal analytic
workflows. Concord provides a query API over
records containing audio, video, transcripts, source-time views,
frames, detections, and tracks. Users construct \emph{Video Relational
Algebra} (VRA) queries through this API. Concord then optimizes the VRA queries by rewriting them to using different media inputs, temporal
extent, or processing granularity.

In sum, this paper makes three contributions:
\begin{itemize}
  \item We define VRA: its
  media model,
  source-time views, deterministic and semantic functions, and operators for various video-related operations (\autoref{sec:vra}).
  
  \item We describe three VRA rewrites: modality substitution,
  cross-modal temporal pushdown, and the algebraic decomposition of whole-video semantic association into
  Detect--Track--Join (\autoref{sec:optimizations}).
  \item We evaluate Concord's query rewrites on real-world videos.
  Transcript-to-video queries improve the cost--quality frontier for
  narrated event localization (\autoref{sec:eval-soccer} and \autoref{sec:eval-lectures}), while Detect--Track--Join improves
  cross-camera F1 from .364 to .813 with no MLLM calls
  (\autoref{sec:eval-cross-camera}).
\end{itemize}

Through our case studies, Concord presents an 
opportunity for future semantic video processing systems, where modality, temporal extent, and processing granularity should be
explicit query choices, subject to event- and entity-coverage constraints.

\section{Background and Related Work}
\label{sec:related}

\paragraph{Video query processing.}
Video DBMSs reduce semantic-inference cost using specialized execution
alternatives. NoScope~\cite{noscope} constructs model cascades and neural
proxies; TASTI~\cite{tasti}, EVA~\cite{eva}, and Seiden~\cite{seiden} use semantic indexes,
materialized inference results, sampling, and temporal propagation.
Spatialyze~\cite{spatialyze} exploits spatial and temporal metadata, while
MIRIS~\cite{miris} integrates query planning with object tracking. These
systems optimize which models, frames, or stored results are evaluated.
Concord builds on this by making media-derived representations composable within a
relational query, enabling alternatives that change the representation,
temporal extent, or record granularity supplied to semantic computation.

\paragraph{Declarative video and multimodal queries.}
VIVA~\cite{viva} is closest to Concord in making relationships between semantic functions
declarative. Its relational hints specify when one registered model may
replace or filter another, including the use of transcript search as a filter
for visual recognition. VOCAL-UDF~\cite{vocaludf} supports compositional video
queries by constructing missing program-based or distilled-model
UDFs. VRA differs by representing aligned transcripts,
temporal clips, detections, and tracks as query-visible values and relations
that preserve source identity and time.

Multimodal systems provide complementary interfaces. ThalamusDB~\cite{thalamusdb} evaluates
natural-language predicates over visual, audio, and textual data; CAESURA~\cite{caesura}
generates executable multimodal queries from natural language; and KathDB~\cite{kathdb}
provides a unified relational interface for multimodal data. Systems such as Palimpzest~\cite{palimpzest}, LOTUS~\cite{lotus}, and DocETL~\cite{docetl,wei2026moar} optimize declarative semantic operators over unstructured data. Concord focuses specifically on algebraic rewrites that exchange or compose source-aligned media representations while preserving the query’s output schema.

\section{Concord and Video Relational Algebra}
\label{sec:vra}

Users express semantic video queries in Concord by writing Video
Relational Algebra (VRA) queries. A VRA query is an expression composed of VRA
operators. Concord may rewrite an initial query into alternative queries that
preserve the query intent and output schema.
\autoref{sec:optimizations} describes these rewrites, which may change the
input modality, temporal extent, or processing granularity.

\subsection{Media Data Model}
\label{sec:data-model}

VRA extends the relational data model with first-class media values. A
relation may contain scalar attributes, nested records, collections, and
media-valued attributes. A video-valued attribute $v$ contains a reference,
such as a local file path or URI, through which VRA operators 
(\autoref{sec:vra-operators}) access the corresponding video content.

Each video is identified by a unique
$\mathit{source\_id}$ that is retained by all derived values and records.
While $v$ provides access to the video content, $\mathit{source\_id}$ identifies its origin, allowing 
independently processed clips and their results to be regrouped by source.

Temporal representations derived from a video use \emph{source time}.
A source-time timestamp is measured on the timeline of the original
video. Temporal information is represented using ordinary
attributes, such as \texttt{start} and \texttt{end} for an interval. Derived
media values, such as clips, retain their source-time intervals, so
clip-relative results can be mapped back to the original video.
Shared source identity and source time let VRA compose media-derived values
and rewrite queries without losing their association with the originating
video or its timeline.

A VRA relation may represent media at different \emph{processing
granularities}: each tuple may correspond, for example, to a complete video, a clip, 
a frame, a detection, or a track. The operators in the following section
transform both the values stored in these tuples and the granularity at which
subsequent computation is performed.

\subsection{VRA Operators}
\label{sec:vra-operators}

A VRA query transforms an input relation into an output relation by composing
VRA operators. Relational operators such as Map, Filter, Reduce, Unnest, and
Join transform and combine relations whose tuples may contain ordinary and
media-valued attributes. Media operators expose common media
transformations such as materialization, transcription, detection, and
tracking. 
\autoref{sec:optimizations} shows how Concord uses these transformations to construct alternative approximate queries.

Media operators have tuple-wise Map semantics: they read designated attributes
from each tuple, store their result in a designated output attribute, and
preserve the remaining attributes. For example,
\[
  \mathsf{View}_{v,s,e\rightarrow v_c}(R)
  \equiv
  \mathsf{Map}_{\mathsf{view}(v,s,e)\rightarrow v_c}(R).
\]
This equivalence specifies the relational behavior of View, while its named
form keeps materialization explicit in VRA queries and rewrites. Transcribe,
Detect, and Track follow the same tuple-wise, attribute-preserving convention.

Relational operators may be parameterized by deterministic or semantic functions. Deterministic functions execute ordinary non-MLLM code, whereas semantic functions are defined by natural-language instructions and a structured output schema and may invoke an MLLM. The operator determines how records are processed, while the function supplies the query-specific computation.

\autoref{tab:concord-operators} defines the VRA operators used in this paper.
In the table, attributes and parameters before $\rightarrow$ are inputs,
and the attribute after $\rightarrow$ stores the result. Operators preserve
attributes not explicitly replaced.
Together, these operators express queries
ranging from semantic classification and aggregation to temporal localization
and cross-video entity association.

\begin{table}[t]
\centering
\caption{Video Relational Algebra operator catalog.}
\label{tab:concord-operators}
\small
\begin{tabular}{@{}>{\raggedright\arraybackslash}p{0.23\columnwidth}
                    >{\raggedright\arraybackslash}p{0.73\columnwidth}@{}}
\toprule
{\bf Operator} & {\bf Semantics} \\
\midrule

\multicolumn{2}{@{}l}{\bf Source operator} \\
$\mathsf{Input}(u)$
& Create a relation from records stored at $u$, including records containing
media values. \\

\addlinespace
\multicolumn{2}{@{}l}{\bf Relational operators} \\
$\mathsf{Map}_{f\rightarrow A}(R)$
& Apply $f$ to each tuple and store result in the field(s) specified by $A$. \\

$\mathsf{Filter}_{p}(R)$
& Retain tuples for which predicate $p$ holds. \\

$\mathsf{Reduce}_{K,g\rightarrow A}(R)$
& Group tuples by $K$, produce one tuple per group, and store the result of
$g$ in attribute $A$. \\

$\mathsf{Unnest}_{A}(R)$
& Expand collection-valued field $A$ into one tuple per element. \\

$\mathsf{Join}_{p,s}(R,S)$
& Combine tuples satisfying predicate $p$ and optionally assign matching
score $s$. \\

$\mathsf{Resolve}_{f}(R)$
& Apply a set-level resolution function $f$ to remove, merge, or select among
conflicting or equivalent records. \\

\addlinespace
\multicolumn{2}{@{}l}{\bf Media operators} \\
$\mathsf{View}_{v,s,e\rightarrow v_c}(R)$
& Materialize interval $[s,e)$ of $v$ and store the resulting clip in $v_c$. \\

$\mathsf{Transcribe}_{v\rightarrow t}(R)$
& Derive source-aligned timestamped text from $v$ and store it in $t$. \\

$\mathsf{Detect}_{v,C\rightarrow D}(R)$
& Detect instances of classes $C$ in $v$ and store the timestamped detections
in $D$. \\

$\mathsf{Track}_{D\rightarrow T}(R)$
& Link detections $D$ within each video source and store the resulting tracks
in $T$. \\

\bottomrule
\end{tabular}
\vspace{-0.18in}
\end{table}

\subsection{Example Queries}
\label{sec:queries}

We now show how to use VRA to write queries.
$Q_1$ and $Q_2$ are temporal event-localization queries that motivate
our cross-modal rewrites (\autoref{sec:modal-sub-rewrite}, \autoref{sec:temp-pushdown-rewrite}).
$Q_3$ is a cross-camera entity-association query that combines semantic processing with
structured media records and motivates our semantic-to-structure rewrite (\autoref{sec:detect-track-join-rewrite}).

\subsubsection{$Q_1$: Soccer goal localization}
\label{sec:ex-soccer}

A sports analyst wants to locate every goal in a collection of soccer
broadcasts. For $Q_1$ and $Q_2$, let
\[
 R_E(source\_id,v,t)
\]
contain one tuple per input video. The attribute $\mathit{source\_id}$
identifies the originating video, $v$ is a media reference through which the
video content can be accessed, and $t$ is its source-aligned timestamped
transcript.

Given an event description $p$, the semantic function
$\mathsf{localize}(v,p)$ returns a collection $E$ of event records represented
as source-time points or intervals. The video-only query applies
this function to $v$ and does not read the transcript $t$:
\begin{equation*}
 \mathsf{Unnest}_{E}\!\left(
   \mathsf{Map}_{\mathsf{localize}(v,p)\rightarrow E}(R_E)
 \right).
\end{equation*}
Map applies the localization function to each input video and stores the
returned collection in $E$ while preserving the input attributes. Unnest
then emits one tuple per event occurrence. For $Q_1$, $p$ specifies that an
actual goal is scored, so the query returns one source-time timestamp for every
goal in each soccer video.

\subsubsection{$Q_2$: Lecture event localization}
\label{sec:ex-lecture}

A student wants the exact portions of a lecture in which a physical
demonstration occurs, excluding explanations, setup, and discussion of the
demonstration. For example, $p$ may request every interval in which a
sustained tone visibly causes a drinking glass to shatter.

$Q_2$ instantiates the localization expression above over lecture videos.
Here, satisfying $p$ requires the co-occurrence of audible and visible
evidence: a passage that discusses the experiment or contains the tone
without the visible shattering does not qualify. The query therefore returns
the source-time intervals containing the requested physical event.

\subsubsection{$Q_3$: Cross-camera vehicle association}
\label{sec:ex-track}
A traffic analyst has synchronized videos from cameras observing overlapping
regions of the same highway. The analyst wants to determine which vehicle observations
across the camera feeds correspond to the same physical vehicle and track
its movement across cameras. The desired output is one cross-camera
trajectory record for each vehicle visible in multiple feeds.
Let
\[
 R_M(\texttt{source\_id},v)
\]
contain one tuple per input camera video. $Q_3$
returns one association record per physical vehicle visible in multiple
cameras:
\[
 (\texttt{vehicle\_id},\texttt{ attributes},
   \texttt{ timeline},\texttt{ match\_score}).
\]
A semantic Reduce--Unnest query places all synchronized input videos in one
group and invokes a semantic function 
$\mathsf{Sem\_Associate}$.
The function discovers vehicles in each video,
determines which observations correspond to the same physical vehicle across
cameras, and returns a collection $A$ of cross-camera association records.
The following VRA query uses this association function to solve the cross-camera joining task:
\begin{equation*}
\mathsf{Unnest}_{A}\!\left(
  \mathsf{Reduce}_{
    \emptyset,\mathsf{Sem\_Associate}\rightarrow A
  }(R_M)
\right).
\end{equation*}
$\mathsf{Sem\_Associate}$ returns a collection $A$ of cross-camera association
records, and $\mathsf{Unnest}$ produces one tuple per associated vehicle. Unlike
$Q_1$ and $Q_2$, this query returns entities and their cross-source
relationships as structured records. Downstream VRA operators can therefore filter, group, join, and process these records as normal relational data.

\section{Optimizing VRA Queries}
\label{sec:optimizations}

VRA exposes three properties of video processing as query choices:
the representation (i.e., modality) supplied to a semantic function, the temporal extent
processed by that function, and the granularity of its intermediate records.
Concord uses these properties to rewrite queries via modality
substitution, temporal candidate pushdown, and operator decomposition to improve query performance.

Each rewrite specifies the properties that the rewritten query must satisfy
and preserves its output schema, but may change result quality by altering the
evidence presented to semantic functions. Quality may degrade when the new
representation omits relevant evidence, or improve when it removes irrelevant
or excessive context, since longer context can reduce model performance even
when the relevant information is present~\cite{du2025context}. We therefore
call these rewrites \emph{approximate} and use $\rightsquigarrow$ to denote
them below.

\subsection{O1: Modality Substitution}
\label{sec:modal-sub-rewrite}

Suppose relation $R$ contains representations $x$ and $y$ derived from the
same media source. We call these representations \emph{source-aligned}: they
share the same $\mathit{source\_id}$ and, when temporal, express timestamps in
source time. Let $f^{x}$ and $f^{y}$ be schema-compatible semantic functions over
their respective representations, parameterized by task description $p$. O1 substitutes one representation for the other:
\[
\mathsf{Map}_{f^{x}(x,p)\rightarrow A}(R)
\;\rightsquigarrow\;
\mathsf{Map}_{f^{y}(y,p)\rightarrow A}(R).
\tag{O1}
\]

The surrounding operators remain unchanged because both implementations
satisfy the same output schema. The rewrite is beneficial when
$y$ is less expensive to process and contains sufficient evidence for
the task. A materialized representation, such as a transcript, can also be
reused across multiple queries.

O1 is not restricted to temporal localization. It can be applied to any
semantic operation admitting schema-compatible implementations over different
modalities, including classification, extraction, and summarization. For $Q_1$ and $Q_2$, O1 replaces the video event-localization function
with a transcript event-localization function. We call the resulting
alternative the \emph{transcript-only query}.

\subsection{O2: Cross-Modal Temporal Candidate Pushdown}
\label{sec:temp-pushdown-rewrite}

Even when a source-aligned representation cannot answer a semantic query
directly, it may identify the source-time regions containing the required
evidence. O2
uses a source-aligned representation to restrict the temporal extent processed by a video
semantic function. Conceptually, it transforms a source-aligned representation into
candidate intervals, materializes those intervals as video clips, evaluates
the clips, and combines their results by source video.

Let $f^{V}(v,p)$ denote a semantic function applied to video $v$ under task
description $p$. A source-time interval is a \emph{temporal witness} for a
result produced by $f^{V}$ when the evidence within that interval is sufficient
to establish the result. O2 applies when the relevant results have bounded
temporal witnesses and results obtained from separate windows can be reconciled
into the function's original output schema. It does not apply when the answer
depends on the entire video, such as determining that an event
never occurs.

\paragraph{Candidate construction.}
Let $R(\mathit{source\_id},v,x)$ contain one tuple per source video, where $x$
is a representation whose timestamps are expressed in the source time of $v$.
The recall-oriented semantic function $\mathsf{cand}(x,p)$ returns a
collection $C$ of candidate intervals in source time. In our prototype,
Concord implements $\mathsf{cand}$ using an MLLM prompt conditioned on $p$. Concord constructs the
materialized candidate clips as follows:
\[
\begin{aligned}
 R_C &=
 \mathsf{Unnest}_{C}\!\left(
   \mathsf{Map}_{
     \mathsf{cand}(x,p)\rightarrow C}(R)
 \right),\\
 R_I &=
 \mathsf{Resolve}_{\mathsf{overlap}}\!\left(
   \mathsf{Map}_{
     \mathsf{window}(c)\rightarrow(s,e)}(R_C)
 \right),\\
 R_W &=
 \mathsf{View}_{v,s,e\rightarrow v_c}(R_I).
\end{aligned}
\]
Unnest produces one tuple per candidate $c$. The deterministic function
$\mathsf{window}$ adds temporal context and returns source-time boundaries
$(s,e)$. Resolve coalesces overlapping windows from the same source, and View
materializes every remaining interval as a standalone clip $v_c$. By its
tuple-wise Map semantics, View preserves the input attributes while adding
$v_c$. Every
tuple in $R_W$ therefore retains its \texttt{source\_id} and source-time
boundaries $(s,e)$.

\paragraph{Rewrite.}
The video semantic function is applied independently to each materialized
clip, and the resulting clip-level records are grouped by source video:
\[
R_A =
\mathsf{Reduce}_{
  \mathit{source\_id},
  \mathsf{reconcile}_{f}\rightarrow A
}\!\left(
  \mathsf{Map}_{
    f^{V}(v_c,p)\rightarrow A_c
  }(R_W)
\right).
\]
The inner Map applies $f^{V}$ to each materialized clip and stores its
clip-level result in $A_c$. Because one source video may produce several
clips, the outer Reduce groups these results by $\mathit{source\_id}$ and
applies $\mathsf{reconcile}_{f}$. The reconciliation function combines the
clip-level results and returns one value $A$ satisfying the output schema
of full-video execution.
O2 is therefore the rewrite
\[
 \mathsf{Map}_{
   f^{V}(v,p)\rightarrow A}(R)
 \;\rightsquigarrow\;
 R_A.
 \tag{O2}
\]
Both sides associate one result $A$ with each \textit{source\_id}; therefore,
the operators following the rewritten expression remain unchanged.

\paragraph{Example: event localization.}
For $Q_1$ and $Q_2$, $x$ is the timestamped transcript, $f^{V}$ is the video
event localizer, and $A$ is the event collection $E$. The candidate function
identifies transcript-grounded regions that may contain the requested event.
The same video localizer used by the video-only query is then applied to each
materialized clip, producing clip-relative event predictions $E_c$. In this
instance, $\mathsf{reconcile}_{f}$ converts each prediction from clip-relative
time to source time and coalesces duplicate predictions. The unchanged
downstream Unnest emits one tuple per event. This rewrite produces the transcript-to-video query evaluated in
\autoref{sec:eval}.

O2 is a cross-modal analogue of selection pushdown: it restricts the input to
a cost-dominant semantic operation using evidence from a source-aligned
representation. Because the retained intervals are generated approximately,
the rewrite is characterized by two quantities. \emph{Candidate recall} is
the fraction of reference results whose temporal witnesses are covered by at
least one retained window. \emph{Selectivity} is the union duration of the
retained windows divided by the duration of the source video. O2 is most effective
when candidate recall is high and selectivity is low.

\subsection{O3: Semantic Reduce--Unnest to Detect--Track--Join}
\label{sec:detect-track-join-rewrite}

For a query that associates the same physical entity across a pair of
synchronized video sources, a semantic Reduce--Unnest query can pass both
videos to one MLLM-backed function. The MLLM must simultaneously discover
entities, maintain their within-video identities, associate entities across
the two sources, and format the resulting associations. Although concise,
this query provides no inspectable intermediate records.

The Detect--Track--Join query decomposes the association into explicit
operators and intermediate relations for detection, tracking, candidate
matching, conflict resolution, and trajectory construction:
\[
\begin{aligned}
R_D &= \mathsf{Detect}_{v, C \rightarrow D}(R),\\
R_T &= \mathsf{Unnest}_{T}\!\left(
  \mathsf{Track}_{D' \rightarrow T}\!\left(
  \mathsf{Map}_{\mathsf{embed}\circ\mathsf{NMS}(D) \rightarrow D'}(R_D)
  \right)
\right),\\
R_J &= \mathsf{Resolve}_{\text{one\_to\_one}}(\mathsf{Join}_{p,s}(R_T, R_T)),\\
R_A &= 
  \mathsf{Map}_{\text{to\_association} \rightarrow assoc}(R_J).
\end{aligned}
\]
where $assoc = (\text{vehicle\_id}, \text{attributes}, \text{timeline}, \text{match\_score})$.

O3 is the rewrite:
\[
\mathsf{Unnest}_{A}\!\left(
  \mathsf{Reduce}_{
    \emptyset,\mathsf{Sem\_Associate}\rightarrow A
  }(R)
\right)
\;\rightsquigarrow\;
R_A.
\tag{O3}
\]

Here, $R$ contains videos from two synchronized sources, $C$ is the set of relevant entity classes, and $R_D$ contains frame-level detections. Within $\mathsf{Map}$, non-maximum suppression removes redundant boxes, while $\mathsf{embed}$ extracts an appearance vector from each entity crop, producing the enriched detection collection $D'$. $\mathsf{Track}$ links detections across frames into tracks $T$, and $\mathsf{Unnest}$ produces the relation $R_T$ with one tuple per within-video track.

$\mathsf{Join}$ self-joins $R_T$. Its predicate $p$ admits only tracks from different sources with compatible motion and 
imposes a canonical ordering on source pairs, thereby removing same-source and symmetric duplicate pairs. The scoring function 
$s$ ranks the remaining candidates using signals such as appearance similarity. Because one track may appear in multiple candidates, $\mathsf{Resolve}_{one\_to\_one}$ greedily processes candidates in descending score order and retains a pair only if neither track has already been matched within that source pair. Finally, $\mathsf{to\_association}$ maps each resolved match to exactly one tuple containing $(\text{vehicle\_id}, \text{attributes}, \text{timeline}, \text{match\_score})$; intermediate track and join fields are not returned.

This rewrite replaces a monolithic MLLM operation with inspectable relations
for detections, tracks, candidate pairs, resolved matches, and associations.
These intermediates expose failure points and enable component-level
optimization while eliminating MLLM calls from the rewritten query. Its
quality depends on \emph{entity coverage}---the fraction of reference
vehicles represented by usable tracks---and \emph{pair coverage}---the
fraction whose correct cross-source match survives candidate pruning and
one-to-one resolution.

\section{Evaluation}
\label{sec:eval}

We have built a prototype of Concord and evaluate the effectiveness of our rewrites using the queries defined above. We evaluate $Q_1$ and $Q_2$ using the video-only query, the
transcript-only query (O1), and the transcript-to-video query (O2).
We evaluate $Q_3$ using the semantic Reduce--Unnest query and the
Detect--Track--Join query (O3). For each event-localization workload, the video-only and transcript-to-video
queries use identical task-specific video-localization instructions and output
schema. Soccer and lectures use different prompts. The shared
video-localization stage receives a complete input video in the video-only
query and transcript-selected materialized clips in the transcript-to-video query.

\paragraph{Datasets.}
\emph{Soccer} contains three SoccerNet matches with English
commentary~\cite{soccernet}, represented as six half-match videos totaling 4.59 hours at 224p, with ten labeled goals. \emph{Lectures} contains three 360p videos 
~\cite{mitocw803}, totaling 3.92 hours and four manually
labeled events. \emph{Highway} uses the first five seconds of two synchronized camera feeds
from the I24V multi-camera highway dataset~\cite{i24v}. We manually
adjudicated a cross-camera reference containing 18 vehicles. Queries and prompts were frozen before evaluation and receive no reference labels.

\paragraph{Runtime methodology.}
We report cold-start end-to-end query execution time. For the video-only
query, this includes transferring each complete input video to the MLLM
provider and performing video event localization. For the transcript-only
query, it includes text-based event localization over the materialized
transcript. For the transcript-to-video query, it includes applying the
candidate function to the materialized transcript to select source-time
windows, materializing the corresponding clips, transferring those clips to
the MLLM provider, and performing video event localization. Transcripts
are treated as materialized inputs generated offline (e.g., from YouTube), so their one-time
construction cost is excluded. No previously uploaded media is reused in the
reported measurements.

\paragraph{Models and implementation.}
We use Gemini 3.1 Flash-Lite~\cite{google2026gemini31flashlite} for all MLLM-backed operators, including video event localization, transcript candidate generation, and the semantic cross-camera association baseline. We report token usage and estimate cost using the provider's pricing at the time of evaluation. The Detect--Track--Join query executes locally on a CPU without GPU acceleration.

\begin{table*}[!t]
\centering
\caption{Event-localization results at the primary thresholds
(Soccer: 30 seconds; Lectures: tIoU $\geq .3$). ``Video'' is the fraction
of source duration reaching the MLLM; ``MLLM Tokens'' include both input
and output tokens.}
\label{tab:eval-main}
\vspace{-0.1in}
\small
\begin{tabular}{@{}llrrrrrrr@{}}
\toprule
workload & query & video (\%) & precision & recall & F1 & MLLM tokens & MLLM cost & query time \\
\midrule
Soccer & video-only
& 100.00 & .750 & .900 & .818
& 1.506M & \$0.378 & 712.3s \\

& transcript-only
& 0.00 & \textbf{.909} & \textbf{1.000} & \textbf{.952}
& \textbf{0.132M} & \textbf{\$0.034} & \textbf{6.3s} \\

& transcript$\rightarrow$video
& 5.32 & \textbf{.909} & \textbf{1.000} & \textbf{.952}
& 0.217M & \$0.057 & 135.8s \\
\midrule
Lectures & video-only & 100.00 & .500 & .500 & .500 & 1.285M &
\$0.322 & 396.2s \\
& transcript-only & 0.00 & .333 & .250 & .286 &
\textbf{0.128M} & \textbf{\$0.032} & \textbf{3.6s} \\
& transcript$\rightarrow$video & 2.47 & \textbf{.800} &
\textbf{1.000} & \textbf{.889} & 0.161M & \$0.041 & 100.5s \\
\bottomrule
\end{tabular}
\vspace{-0.1in}
\end{table*}

\begin{table*}[!t]
\centering
\caption{Highway cross-camera join over two five-second clips and 18
reference vehicles.}
\vspace{-.1in}
\label{tab:eval-cross-camera}
\small
\begin{tabular}{@{}lrrrrrrrrrr@{}}
\toprule
query & preds & TP/FP/FN & precision & recall & F1 &
timeline IoU & attribute exact & wall time (s) & MLLM calls & MLLM tokens \\
\midrule
semantic Reduce$\rightarrow$Unnest
& 4 & 4/0/14 & \textbf{1.000} & .222 & .364
& .576 & \textbf{.583} & \textbf{52.724} & 1 & 1{,}515 \\
Detect$\rightarrow$Track$\rightarrow$Join
& 14 & 13/1/5 & .929 & \textbf{.722} & \textbf{.813}
& \textbf{.762} & .538 & 99.559 & \textbf{0} & \textbf{0} \\
\bottomrule
\end{tabular}
\vspace{-0.1in}
\end{table*}

\subsection{Soccer Goal Localization}
\label{sec:eval-soccer}

Predictions are matched one-to-one with reference
goals. A prediction is correct when its absolute timestamp error
relative to the matched reference goal is at most 30 seconds. We additionally
evaluate tolerances of 5, 10, and 60 seconds to measure sensitivity to temporal
precision.

\autoref{tab:eval-main} reports the results. The transcript candidate
function retains 879 of 16,522 seconds of source video, corresponding to
5.32\% selectivity. Its 13 candidate windows cover all ten reference goals,
yielding 100\% candidate recall.

At the primary 30-second tolerance, the transcript-to-video query improves F1
from .818 for video-only execution to .952. The transcript-only query also
achieves .952 F1 at this tolerance, but produces less precise timestamps:
among matched goals, its mean and median absolute errors are 4.274 and 2.011
seconds, compared with .348 and .181 seconds for transcript-to-video. The F1 results at tighter tolerances confirm this difference. At tolerances of 5, 10, 30, and
60 seconds, transcript-to-video achieves .952 F1 throughout, whereas
transcript-only achieves .762, .857, .952, and .952, respectively;
video-only remains at .818 across all four tolerances. The video stage
therefore refines transcript-derived candidate regions into more precise goal
timestamps.

Relative to video-only execution, the transcript-to-video query reduces total
MLLM tokens by 85.6\%, estimated MLLM cost by 85.0\%, and cold-start wall time
by 80.9\%, from 712.3 to 135.8 seconds. Thus, after temporal pushdown is coupled
with clip materialization, the rewrite improves both model cost and
end-to-end latency while preserving candidate recall.

\subsection{Lecture Event Localization}
\label{sec:eval-lectures}

Video is presented to the MLLM at one frame per
second. Predicted intervals are matched one-to-one with reference intervals
using temporal intersection over union (tIoU), defined as the duration of
their intersection divided by the duration of their union. A prediction is
counted as correct at the primary threshold when its tIoU with the matched
reference interval is at least $.3$. We additionally evaluate thresholds of
$.1$ and $.5$ to measure sensitivity to temporal boundary accuracy.

\autoref{tab:eval-main} reports the results at the primary tIoU threshold.
The transcript candidate function retains 349 of 14,115 seconds of source
video, i.e., 2.47\% selectivity. Its three candidate windows fully
cover all four reference events, yielding 100\% candidate recall.

At tIoU $\geq .3$, the transcript-to-video query improves F1 from .500 for
video-only execution to .889, with precision .800 and recall 1.000. Relative
to video-only execution, it reduces total MLLM tokens by 87.4\%, estimated MLLM
cost by 87.2\%, and cold-start query execution time by 74.6\%, from 396.2 to
100.5 seconds.

The improvement persists across temporal-overlap thresholds.
At tIoU thresholds $.1$, $.3$, and $.5$, transcript-to-video achieves F1
scores of $.889$, $.889$, and $.444$, respectively. The corresponding
video-only scores are $.750$, $.500$, and $.250$, while the transcript-only
scores are $.857$, $.286$, and $.000$. Video refinement
therefore improves temporal localization rather than merely detecting whether
the requested demonstration is present.

The transcript-only query reaches only .286 F1 at the primary threshold
because transcript evidence often identifies a broad semantic neighborhood
rather than the number and precise boundaries of the visible events. For the
query that localizes successful soap-bubble demonstrations, one
transcript-only prediction spans approximately 57 seconds and covers both
six-second reference events instead of returning a separate interval for each.

This behavior differs from commentary displacement in soccer. A soccer goal
is a point event whose verbal description may precede or follow the visible
goal. In the lecture workload, the transcript identifies the relevant
demonstration but may not determine its visual boundaries or distinguish
multiple occurrences within the same discussion interval. Transcript
candidate localization and video event localization therefore perform
complementary roles in the transcript-to-video query.

\subsection{Cross-Camera Vehicle Trajectories}
\label{sec:eval-cross-camera}

The semantic Reduce--Unnest baseline sends both clips to Gemini 3.1 Flash-Lite in one
MLLM-backed Reduce and unnests the returned cross-camera trajectories. The
Detect--Track--Join query uses YOLOE detections~\cite{wang2025yoloe}, motion-aware tracking,
ResNet50 appearance embeddings~\cite{he2016resnet}, pair pruning, and greedy one-to-one matching.
We score each prediction--reference pair as the fixed weighted sum of
$0.7\,\mathsf{tIoU}+0.3\,\mathsf{Attr}$ (coefficients chosen \emph{a priori}
to favor temporal overlap over noisy cross-camera attributes; not tuned on
the reported runs), then assign matches greedily one-to-one.
Here $\mathsf{tIoU}$ is the temporal IoU from earlier, averaged over shared
cameras, and $\mathsf{Attr}$ is soft agreement on class, color, and subtype.
A pair trajectory is accepted if this score is at least $0.70$; the Join itself keeps
candidate pairs with appearance cosine similarity at least $0.40$.

\autoref{tab:eval-cross-camera} summarizes the cross-camera association results.
The semantic Reduce--Unnest baseline is precise for the four trajectories it
returns, but its recall is .222. The Detect--Track--Join query produces 14
trajectories, matches 13 reference vehicles, and raises F1 from .364 to .813
while improving mean timeline IoU from .576 to .762. It also eliminates MLLM
calls by replacing whole-clip inference with explicit Detect, Track, and Join operators.

On these five-second clips, however, the Detect--Track--Join query is slower
than the semantic Reduce--Unnest query ($99.6$\,s vs.\ $52.7$\,s): the
rewrite trades MLLM usage for higher coverage, not end-to-end latency on this
short sample.

\balance
\section{Conclusion}

We described Concord, a multimodal video query processing system built on VRA over video inputs.
By exposing different modalities within a single algebra,
VRA allows users to express queries and also Concord to optimize them.
Our experiments show that Concord's optimizations can significantly reduce cost and query
execution time while improving accuracy, demonstrating their promise for
semantic video query processing.

\bibliographystyle{ACM-Reference-Format}
\bibliography{refs}

\end{document}